\documentclass[%
 reprint,
 amsmath,amssymb,
 aps,
prl,
]{revtex4-2}

\usepackage{graphicx}	
\usepackage[dvipsnames]{xcolor}
\usepackage{gensymb}

\usepackage{natbib}
\usepackage[colorlinks=true, citecolor=RoyalBlue,urlcolor=RoyalBlue]{hyperref}
\newcommand{\Pe}{{\rm Pe}}
\newcommand{\Vext}{V_{\rm ext}}
\newcommand{\epsw}{\varepsilon_w}
\newcommand{\elpar}{\ell_{\parallel}}
\newcommand{\elprp}{\ell_{\perp}}

\begin{document}

\title{Wetting transition of active Brownian particles on a thin membrane}
\author{Francesco Turci}
\email[Corresponding author: ]{f.turci@bristol.ac.uk}
\author{Nigel B.\ Wilding}
\affiliation{H.H.Wills Physics Laboratory, Royal Fort, Bristol, BS8 1TL, U.K.}

\begin{abstract}
 We study non-equilibrium analogues of surface phase transitions in a minimal model of active particles in contact with a purely repulsive potential barrier that mimics a thin porous membrane. Under conditions of bulk motility-induced phase separation, the interaction strength $\epsw$ of the barrier controls the affinity of the dense phase for the barrier region. We uncover clear signatures of a wetting phase transition as $\epsw$ is varied.  In common with its equilibrium counterpart, the character of this transition depends on the system dimensionality{: a continuous transition with large density fluctuations and gas bubbles is uncovered in 2d while 3d systems exhibit a sharp transition absent of large correlations.}  
 \end{abstract}

\maketitle


The motion of natural microswimmers such as bacteria and algae is strongly influenced by their interactions with confining walls, interfaces and barriers \cite{elgeti2015,bechinger2016}. On surfaces, they can form biofilms, which result from the motion, growth and death of the swimmers in the presence of repulsive and hydrodynamic interactions \cite{lauga2006,even2017,hartmann2019}.
This accumulation of dense aggregates in self-propelled, non-equilibrium active systems in contact with walls is reminiscent of the phenomenology of wetting that occurs in equilibrium fluids. Wetting is a surface phase transition that can occur when a fluid at bulk liquid-gas coexistence is brought into contact with an attractive wall. The relevant behaviour is characterised by the macroscopic contact angle that a sessile liquid drop makes with the wall. In this setting, mechanical and thermal equilibrium are equivalent and the stability of the droplet is captured by Young's equation relating gas-liquid, wall-gas, wall-liquid surface tensions and the contact angle, $\gamma_{\rm lg}\cos\theta=\gamma_{\rm wg}-\gamma_{\rm wl}$. For $0\leq\cos(\theta)<1$ the system is said to be partially wet, with a wetting transition occurring as $\cos(\theta)\to 1$. For $-1\leq\cos(\theta)<0$ the system is partially dry, with a drying transition (the limit of extreme hydrophobicity in which the wall favors the gas phase) occurring as $\cos(\theta)\to -1$. The attractive strength of wall-fluid interactions determines $\gamma_{\rm wg}$ and $\gamma_{\rm wl}$ and hence $\theta$. Wetting and drying transitions can be first order or critical (i.e. continuous) depending on the properties of the wall-fluid interactions and the system dimensionality \cite{binder2003, bonn2009,wu2016a,evans2017,evans2019}.

 Self-propelled particles belong to a class of active matter systems that manifest a type of gas-liquid phase coexistence which is termed motility induced phase separation (MIPS). In contrast to fluid phase coexistence in equilibrium systems, MIPS can occur even in the absence of attractive particle-particle interactions \cite{cates2015motility}. MIPS emerges from the increased persistence of motion of self-propelled particles coupled to steric interactions. It has been studied in detail in both two-dimensional (2d) and three-dimensional (3d) systems via a minimal model, so-called \textit{active Brownian particles} (APBs). These are characterised by purely repulsive, isotropic interactions and an internal driving force  (or velocity) whose direction constantly diffuses on the unit sphere \cite{siebert2018,digregorio2018,wysocki2014,turci2021}. ABPs in the bulk thus display qualitative similarities with equilibrium systems, which have motivated mappings to effective interactions and free energies \cite{solon2018a,speck2021}.

Although the existence of bulk phase separation in active matter seems settled, there is no clear consensus on the nature of interfacial properties and surface phase transitions. In the most extensively studied case of 2d systems, the interfaces that form between coexisting bulk phases exhibit far greater fluctuations than seen in equilibrium fluids, with a propensity to form defects or bubbles of the less dense phase \cite{caporusso2020}. A suitable definition of surface tension between coexisting phases is still contentious \cite{bialke2015,hermann2019a,speck2020,omar2020}. 

{Quite generally one observes accumulation of active particles at a purely repulsive `hard' wall or impenetrable obstacles \cite{bechinger2016,das2018}. This has led to numerical searches for surface phase transitions in 2d models of active matter  \cite{sepulveda2017,sandor2017,sepulveda2018,neta2021a}. For a 2d lattice gas, hard walls lead to a completely wet state having $\theta=0$ \cite{neta2021a}. This contrasts with equilibrium liquids, for which a hard wall promotes a dry state \cite{evans2019} where the vapor phase is in contact with the wall. The question arises whether for active matter one can have a wall that is partially wet or partially/completely dry and whether one can observe transitions between these states. As we show in this Letter, to do so one must look beyond the case of hard walls and consider the effects of a finite potential barrier (representative of a thin permeable membrane).} We find that even though active particles respond to such a barrier in a manner completely different to their equilibrium counterparts, a clear signature of a wetting transition can nevertheless be identified. {In 2d, the transition from a partially wet to a completely wet state appears continuous and is accompanied by large density fluctuations and bubbles. By comparison in 3d the transition is much sharper and exhibits no discernible large length scale correlations either parallel or orthogonal to the barrier.}

Our model system comprises purely repulsive spherical active Brownian particles, interacting via the Weeks-Chandler-Anderson (WCA) potential with energy and length scales $\epsilon$ and $\sigma$ - see supplementary material (SM)~\cite{supplementary}. Following previous studies \cite{stenhammar2014,turci2021}, we take the number density $\rho=N/V$ and the P\'eclet number ${\rm Pe}=v_0/\sigma D_r$ as control parameters, where $N$ is the number of particles, $V$ the total volume in units of $\sigma^d$ (with $\sigma$ the particle diameter), $v_0$ is the swimming speed and $D_r$ is the rotational diffusion constant,  coupled to the translational diffusivity by $D_r=3D_t/\sigma^2$, which defines an intrinsic persistent-motion timescale $\tau_R=1/D_R$. This standard setting gives rise to liquid-gas MIPS both in 2d and 3d, with the difference that in the latter case this is metastable with respect to crystal-gas phase separation \cite{turci2021,omar2021}. 

For surface phase transitions to occur, it is necessary for the system to undergo MIPS. Based on the known phase behaviour - see SM~\cite{supplementary}-  we set $\Pe=50$ for 2d, and $\Pe=60$ for 3d, values that are well in excess of the respective critical points $\Pe^{\ast}\approx 25,36$  \cite{stenhammar2014,turci2021}.  Bulk phase separation occurs into a low density (gas-like) region and a high density (liquid-like) region with coexistence densities $\rho_{LD}$ and $\rho_{HD}$.  We fix $\rho=0.5$ in 2d and $\rho=0.75$ in 3d, values that are somewhat smaller than the coexistence diameter $(\rho_{\rm LD}+\rho_{\rm HD})/2$, resulting in similar volumes of each phase within the system. We consider a rectangular {periodic} simulation box having dimensions $L_x>L_y$. {Similarly to an equilibrium simulation in the constant-NVT ensemble, our ABP system with constant-N,V,{\rm Pe} exhibits a liquid slab configuration that spans the system in the $y$-direction.} 

To this system we add a localised external potential, $\Vext(x)$,  a cosine hump centred on $x=0$ that depends only on the $x$ coordinate: $\Vext(x) = \epsw \left[\cos(\pi x/d)+1\right]H(d-x)H(x+d)$, with $H(x)$ the Heaviside function. This form ensures that the repulsive force goes to zero smoothly at a distance $x=d$ from the barrier. We set $d=\sigma$, resulting in a thin, localised barrier whose size is comparable to the particle diameter. The sole barrier parameter is therefore $\epsw$ which we express in units of $\epsilon$, and which controls the repulsive barrier strength: letting $\epsw\to\infty$ yields an impenetrable wall, while $\epsw\rightarrow 0$ results in the free active diffusion of particles on the torus. Intermediate values can be thought of as representing a thin porous membrane with non-zero crossing probability~\cite{das2020}. {For a liquid slab arrangement, isolated liquid droplets exhibiting a contact angle are absent. In analogy with wetting phenomenology in constant-NVT simulations of equilibrium fluids~\cite{vanleeuwen1989,nijmeijer1990} in a slit geometry, we expect that changes in the contact angle corresponding to surface phase transitions manifest as changes in the affinity of the liquid slab for the barrier.}

\begin{figure}[!t]
\centering
  \includegraphics[width=\columnwidth]{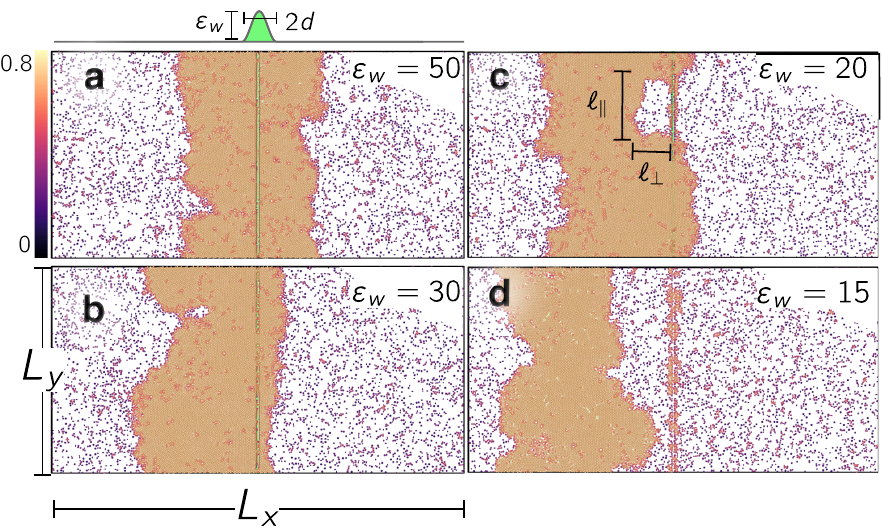}
  \caption{(a-d) Snapshots of the 2d system for decreasing values of the barrier strength $\epsw$, as described in the text. Particles are color-coded by their local density {(see colorbar)}. The geometrical parameters are indicated and a sketch of the applied cosine energy barrier is shown.}
  \label{fig:sym-asym}
\end{figure}

Figure~\ref{fig:sym-asym} shows the qualitative behaviour that a change in the repulsive barrier height induces in the two phase separated system. At sufficiently large $\epsw$ the well-established \cite{bechinger2016} phenomenon of slowing down and entrapment due to impenetrable walls is observed: the dense liquid phase is favoured and becomes localised at the barrier \footnote{Extremely repulsive barriers ($\epsw>100$) push this mechanism further, promoting the formation of ordered layers reminiscent of the bulk ordered hexagonal phase \cite{digregorio2018}.}. At high $\epsw$, Fig.~\ref{fig:sym-asym}(a), this localisation is symmetric in the sense that the steady state exhibits approximately equal-sized liquid layers on both sides of the barrier, regardless of the initial conditions. Accordingly, both sides of the barrier are wet. We quantify the degree of asymmetry of the instantaneous density profile $\rho(x,t)$ with respect to the barrier location via the quantity~\cite{supplementary}

\begin{equation}
    \mathcal{A}(t) =\left| \frac{\int_0^{L_x/2}\rho(x,t)dx-\int_{-L_x/2}^{0}\rho(x,t)dx}{(\rho-\rho_{\rm LD})L_{x} }\right|.
\end{equation}
In the steady state, the average $\overline{\mathcal{A}(t)}$ (over time and distinct initial conditions) provides a measure of the typical asymmetry of the liquid region with respect to $x=0$.

\begin{figure}[t]
\centering
  \includegraphics[]{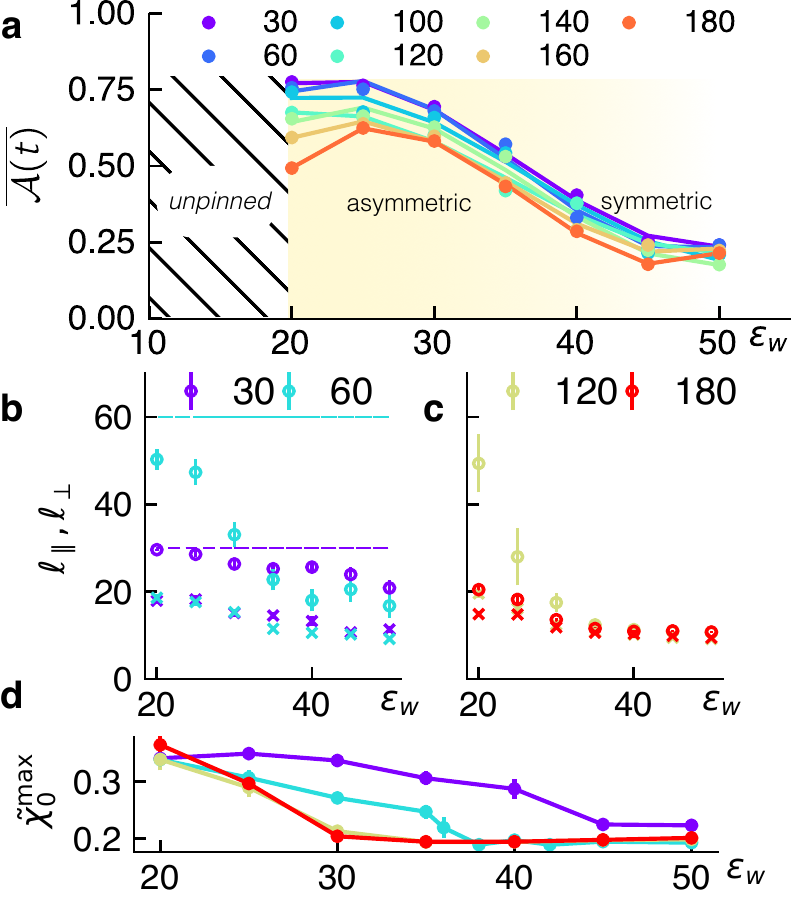}
    \caption{Surface phase behavior in 2d. (a) The density profile asymmetry as the order parameter $\overline{\mathcal{A}(t)}$. (b-c) Top 15 percentile of the parallel ($\circ$) and perpendicular ($\times$) bubble sizes in units of $\sigma$ for various $L_y$  with $L_x=120\sigma$. Horizontal lines in (b) indicate the corresponding value of $L_y$: detachment occurs at $\elpar\simeq L_y$. (d) The maximum value of the local compressibility, for different sizes $L_y$ (colors as in (b,c)).   }
  \label{fig:asym2d}
\end{figure}

As the repulsive strength $\epsw$ of the barrier is lowered, we observe that in the steady state the high density phase migrates such that it is no longer symmetrically localised at the barrier, but instead occupies one side or the other with equal probability, Fig.~\ref{fig:sym-asym}(b) \footnote{Long-time simulations show that rare crossing events do occur with a frequency that depends on the box size and which diminishes with increasing $N$-see SM~\cite{supplementary}}. { It is instructive to compare this symmetry change with that occurring when an equilibrium fluid studied within the constant-NVT ensemble undergoes a wetting transition.  There one observes \cite{vanleeuwen1989,nijmeijer1990}} a transition from symmetric to asymmetric density profiles strongly reminiscent of the phenomenon that we have identified. The symmetry change follows from Young's equation  {as discussed further in the SM~\cite{supplementary}}: at the wetting point $\cos(\theta)=1$ and thus $\gamma_{lg}=\gamma_{wg}-\gamma_{wl}$. This implies that the free energy of an asymmetric profile with three interfaces and surface tension $\gamma_{\mathrm{tot}}=\gamma_{\mathrm{wg}}+\gamma_{\mathrm{wl}}+\gamma_{\mathrm{lg}}$ is equal {at the transition} to that of a symmetric one with $\gamma_{\mathrm{tot}}=2\gamma_{\mathrm{wl}}+2\gamma_{\mathrm{lg}}$.  
Of course for active matter there is as yet no agreed definition of a surface tension. Nevertheless the existence of interfaces implies that an analogous quantity should exist and based on arguments of mechanical equilibrium, it must also obey Young's equation. Hence, we interpret the change in the symmetry of the density distribution in the active systems as a wetting transition. Figure \ref{fig:asym2d}(a) quantifies the time averaged asymmetry $\overline{\mathcal{A}(t)}$ in 2d with respect to variation in $\epsw$, which serves as an order parameter for the wetting transition. We note a systematic but weak finite-size dependence of the continuous transition from symmetric (high $\epsw$) to asymmetric (low $\epsw$) profiles, with larger systems displaying a smoother transition.


As we reduce the barrier strength even further, Fig.~\ref{fig:sym-asym}(c), more features emerge. While the barrier continues to promote the accumulation of the high-density phase on one side, large density fluctuations occur within the liquid-like phase close to the barrier. In 2d systems, MIPS is characterised by droplets and bubbles within the bulk phases \cite{caporusso2020}. However, here we observe the formation of large anisotropic bubbles which are localised along the barrier. The size of the bubbles depends on $\epsw$: for weak barriers, $\epsw\leq 20$, they can grow to such an extent that a gas layer spans the length of the barrier causing the liquid slab to detach, Fig.~\ref{fig:sym-asym}(d). 

The behaviour at small $\epsw$ is similar to that occurring in equilibrium fluids at a critical drying transition \cite{evans2016,evans2017}. To analyse it further, we define two characteristic  length scales $\elpar,\elprp$ which measure the dimensions of bubbles parallel to, and perpendicular to the barrier respectively. These quantities were measured over very many steady state configurations allowing accumulation of their probability distributions as described in the SM~\cite{supplementary}. The distributions of $\elpar,\elprp$ depend on $\epsw$. While a large number of small bubbles having $\elpar\sim\elprp\approx 10\sigma$ are present for all $\epsw$ -- reflecting the bubble-forming propensity of the bulk liquid -- for small $\epsw$ significant large deviations can be observed in the form of rare very large bubbles. To focus on these extreme values we restrict attention to those bubbles who sizes are in the highest 85th percentile of the distribution. With reducing $\epsw$, we find that $\elpar$ and $\elprp$ for these bubbles grow continuously as shown in Fig.~\ref{fig:asym2d}(b,c). Notably these extremal bubbles are predominately located near the barrier and have $\elpar>\elprp$ i.e. the bubbles  are `flat', cf. Fig.~\ref{fig:sym-asym}(c). The largest bubbles can be very large indeed at small $\epsw$ and this effect is more pronounced for systems with a smaller transverse length $L_y$: in fact, the formation of bubbles that span the system parallel to the barrier was observed for systems of sizes up to $L_y=60\sigma$, though not for larger systems. The detachment arising from a spanning gas layer is reminiscent of the finite-size effect `premature drying'~\cite{evans2017} seen in equilibrium fluids, in which a growing correlation length parallel to the attractive wall attains the transverse system size and causes the liquid layer to unbind, thereby preempting the true surface critical behaviour.
 
The parallel correlation length is intimately linked to the local compressibility which is extracted from the density profiles $\left\langle\left(\delta N(x)\right)^{2}\right\rangle=\left\langle\left(N(x)-\left\langle N(x)\right\rangle\right)^{2}\right\rangle$, where $N(x)$ is the profile of the number of particles along the $x$ direction. Following Ref \cite{phys2014}, we define a scaled compressibility profile $\chi(x)= \left\langle\left(\delta N(x)\right)^{2}\right\rangle/\langle N(x)\rangle$, with the average corresponding to a time average over the steady state. This exhibits a peak in the vicinity of the barrier, whose height $\tilde{\chi}^{\rm max}_{0}$ is a measure of the strength of local density fluctuations, see Fig.~\ref{fig:asym2d}(d) and the SM~\cite{supplementary}. Similarly to $\elpar$, this quantity increases continuously as $\epsw$ is reduced although the onset value of the increase depends on $L_y$. We also find that the smaller systems exhibit typically larger $\tilde{\chi}^{\rm max}_{0}$ even at large $\epsw$.

 \begin{figure}[t]
\centering
  \includegraphics[]{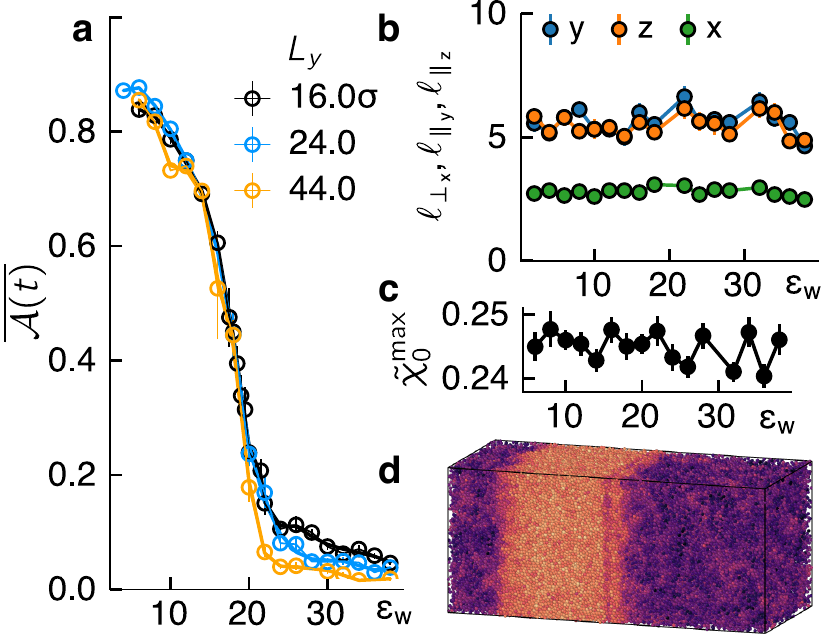}
  \caption{Surface phase behavior in 3d. (a) Asymmetry order parameter $\overline{\mathcal{A}(t)}$ for various $\epsw$ and orthogonal system size $L_y=L_z$. The wetting point is $\epsw^\times\approx 18(1)$. (b) Dimensions $\elpar,\elprp$ of low density bubbles within the high density region for $L_y=L_z=32\sigma$. (c)  The maximum value of the local compressibility as a function of $\epsw$. (d) A snapshot at $\epsw=14$ with $L_y=L_z=44\sigma$ showing the system in the asymmetric state with particles color-coded by their local density; the central density depletion is due to the repulsive barrier.}
  \label{fig:3d}
\end{figure}

\begin{figure}[t!]
\centering
  \includegraphics{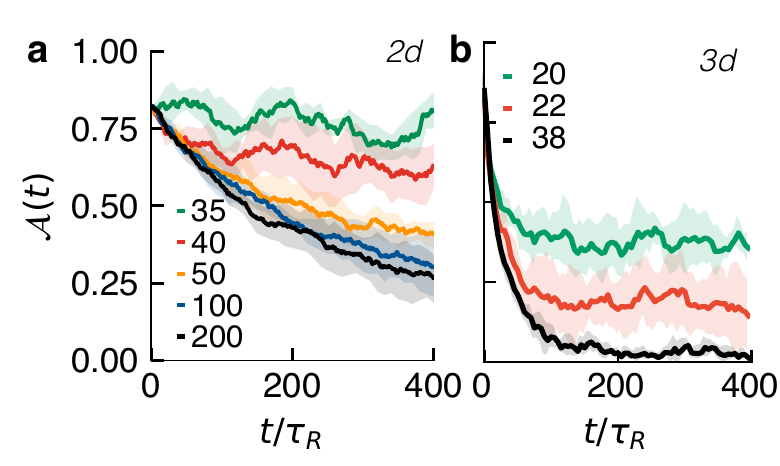}
  \caption{Time evolution of $\mathcal{A}(t)$ following `quenches' from the asymmetric (partially wet) phase to the symmetric (wet) phase in 2d (a) and 3d (b). Systems are prepared in an asymmetric state at $\varepsilon_w^{2d}=25$ and  $\varepsilon_w^{3d}=6$ and quenched instantaneously to the  $\epsw^{\prime}$ values in the key. {Lines are averages over 6 independent trajectories and shaded areas 1 standard deviation fluctuations. In 2d, $L_x=240\sigma,L_y=120\sigma,N=14\ 400$, in 3d $Lx=100\sigma,L_y=L_z=24\sigma,N=43\ 200$}.}
  \label{fig:quench}
\end{figure}

Dimensionality can greatly affect the character of surface phase transitions.  For example, equilibrium wetting is continuous in 2d~\cite{abraham1980,wu2016a} and discontinuous in 3d~\cite{evans2019}. To investigate dimensionality effects, we have studied a 3d system of ABPs in an arrangement similar to that of the 2d system described above. Initially uniform systems were permitted to phase separate  in the presence of barriers of various repulsive strengths $\epsw$. Once the steady state was reached, we tracked local density fluctuations and density profiles. Our results reveal several similarities with 2d: (i) a wetting transition from symmetric to asymmetric states as $\epsw$ is reduced from large to small values, see Fig.~\ref{fig:3d}(a,d); (ii) weak dependence of the transition point on the lateral system size, Fig.~\ref{fig:3d}(a); (iii) depinning of the liquid slab from the barrier for small but finite $\epsw\lesssim 1$. However, there are key differences with the 2d case: (1) the asymmetric-to-symmetric transition is sharp, and becomes sharper with increasing system sizes, see Fig.~\ref{fig:3d}(a), and (2) regardless of the repulsive barrier height, {the local compressibility maximum $\tilde{\chi}^{\rm max}_{0}$, Fig.~\ref{fig:3d}(c), at the barrier does not vary significantly, indicating absence of a growing parallel correlation length (see SM\cite{supplementary}). Measurements of typical bubble sizes across the entire range of $\epsw$ show no evidence of a developing large length scale that would signal critical drying Fig.~\ref{fig:3d}(b). }

 The sharpness of the 3d wetting transition {and the absence of large-scale fluctuations at the barrier as measured by $\chi(x)$} suggests first order behaviour and this led us to investigate whether metastability is associated with the transition and to compare with the 2d case.  We prepared systems in the weakly bound asymmetric state at low $\epsw$ and then implemented an instantaneous `quench' to various larger $\epsw^{\prime}$ above the approximate wetting point $\epsw^{\times}$, where  $\epsw^{\times}\approx 30$\:(2d) and  $\epsw^{\times}\approx 18$\:(3d). The associated time evolution of  $\mathcal{A}(t)$ over an interval of $400\tau_R$ is shown in Fig.~\ref{fig:quench} for several values of $\epsw^{\prime}>\epsw^{\times}$ {for an ensemble of 6 trajectories for each condition}. In the 3d case and for quenches just beyond the transition point, there is an initial rapid relaxation which plateaus out for the duration of the simulation while - for higher barrier strength - the system attain the symmetric state for $t<200\tau_R$. By contrast, in 2d,  the decay of asymmetry accelerates only very gradually with increasing barrier strength, so much so that even for quenches to $\epsw^{\prime}\approx 8\epsw^{\times}$ the symmetric state is not attained within the observation time. {These findings suggest that the transition in 3d occurs via fast local mechanisms and exhibits signs of metastability (with respect to the completely wet state). These features are reminiscent of a first order phase transition in equilibrium systems. By contrast, in 2d the slow dynamics indicates that the transition entails relaxation on large length scales.} {Further elucidation of the detailed properties and order of the transitions may require deployment of finite-size scaling techniques~\cite{evans2017}-- a task for future work.}

In conclusion, we have investigated surface phase behaviour of ABPs in contact with a repulsive barrier that mimics the effects of a thin porous membrane. Our work goes beyond previous studies of active matter at impenetrable walls \cite{kaiser2012,ni2015,levis2017,sepulveda2017,sepulveda2018,neta2021a} and yields clear evidence of a wetting transition and establishes its character. This transition emerges from a previously unidentified mechanism: the competition between the density depletion induced by a finite repulsive barrier and activity-driven aggregation on obstacles. Such a mechanism is completely distinct to that which drives wetting and drying transitions in equilibrium fluids where wall-fluid attraction is necessary~\cite{evans2019}. More broadly, our work suggests that established concepts and language developed in the study of surface phase transitions in equilibrium liquids may be useful in elucidating the interfacial properties of active matter. For instance, accurate knowledge of the location of the wetting transition could allow tests of relationships between gas-liquid, wall-gas, and wall-liquid surface tensions in active matter. Our results thus open up new avenues of theoretical enquiry, and are also amenable to experimental tests, e.g. with self-propelled colloids \cite{dreyfus2005,palacci2010,bricard2013} or elementary robots \cite{sen2009, dauchot2019, alhafnawi2021}.


\begin{acknowledgments}
The authors thank R. Evans for insightful conversations and critical reading of the manuscript. This work was carried out using the computational facilities of the Advanced Computing Research Centre, University of Bristol.
\end{acknowledgments}


%

\end{document}